\documentclass[prl,twocolumn,superscriptaddress,showpacs,nobalancelastpage]{revtex4}

\usepackage[pdftex]{graphicx}
\usepackage{float}
\usepackage{braket}
\usepackage{amsmath}

\newcommand{\SI}[2]{\ensuremath{#1\,\mathrm{#2}}}

\begin{document}

\title{Quantum manipulation of two-electron spin states in metastable double quantum dots}

\author{Benoit Bertrand}
\affiliation{Univ. Grenoble Alpes, Inst NEEL, F-38042 Grenoble, France}
\affiliation{CNRS, Inst NEEL, F-38042 Grenoble, France}

\author{Hanno Flentje}
\affiliation{Univ. Grenoble Alpes, Inst NEEL, F-38042 Grenoble, France}
\affiliation{CNRS, Inst NEEL, F-38042 Grenoble, France}

\author{Shintaro Takada}
\affiliation{Univ. Grenoble Alpes, Inst NEEL, F-38042 Grenoble, France}
\affiliation{CNRS, Inst NEEL, F-38042 Grenoble, France}
\affiliation{Department of Applied Physics, University of Tokyo, Tokyo, 113-8656, Japan.}

\author{Michihisa Yamamoto}
\affiliation{Department of Applied Physics, University of Tokyo, Tokyo, 113-8656, Japan.}
\affiliation{PRESTO-JST, Kawaguchi-shi, Saitama 331-0012, Japan.}

\author{Seigo Tarucha}
\affiliation{Department of Applied Physics, University of Tokyo, Tokyo, 113-8656, Japan.}
\affiliation{RIKEN Center for Emergent Matter Science (CEMS), 2-1 Hirosawa, Wako-Shi, Saitama 31-0198, Japan.}

\author{Arne Ludwig}
\affiliation{Lehrstuhl f\"ur Angewandte Festk\"orperphysik, Ruhr-Universit\"at Bochum, Universit\"atsstrasse 150, 44780 Bochum, Germany.}

\author{Andreas D. Wieck}
\affiliation{Lehrstuhl f\"ur Angewandte Festk\"orperphysik, Ruhr-Universit\"at Bochum, Universit\"atsstrasse 150, 44780 Bochum, Germany.}

\author{Christopher B\"auerle}
\affiliation{Univ. Grenoble Alpes, Inst NEEL, F-38042 Grenoble, France}
\affiliation{CNRS, Inst NEEL, F-38042 Grenoble, France}

\author{Tristan Meunier}
\affiliation{Univ. Grenoble Alpes, Inst NEEL, F-38042 Grenoble, France}
\affiliation{CNRS, Inst NEEL, F-38042 Grenoble, France}

\date{\today}

\begin{abstract}
We studied experimentally the dynamics of the exchange interaction between two antiparallel electron spins in a so-called metastable double quantum dot where coupling to the electron reservoirs can be ignored. We demonstrate that the level of control of such a double dot is higher than in conventional double dots. In particular, it allows to couple coherently two electron spins in an efficient manner following a scheme initially proposed by Loss and DiVincenzo \citep{loss-1998_PhysRevA.57.120}. The present study demonstrates that metastable quantum dots are a possible route to increase the number of coherently coupled quantum dots.
\end{abstract}

\pacs{03.65.w, 03.67.Mn, 42.50.Dv}

\maketitle  

An important stream of research is nowadays to develop tools to operate and control quantum nanocircuits at the single electron level. The coherent manipulation of the spin of an electron trapped in a quantum dot is now well established \cite{Hanson07-1, shulman2012demonstration, veldhorst2014addressable}. Some of the most advanced spin manipulation schemes take advantage of the important control of quantum dot systems defined with lateral gates in a two dimensional electron gas (2DEG). However, the possibility to exchange electrons between the dot and the reservoir reduces the available tuning parameter space and renders the manipulation of multidots almost intractable. Isolating the electron from the reservoirs \cite{Rushforth2004,Johnson2005a} could therefore not only restore the full tunability of the dot system and simplify the control of multidot systems but can also remove parasitic effects occuring during electron spin manipulation such as photon assisted tunneling \cite{koppens-nature-2006}.

Here we demonstrate that coupled quantum dots can be defined and well controlled in a metastable configuration above the Fermi energy, where the coupling to the electron reservoir can be ignored. The tunnel coupling between the dots can be easily tuned over several orders of magnitude while keeping the number of electrons in the dots constant. This allows us to perform controlled exchange oscillations between two electron spins by changing only the tunnel coupling between the dots as initially proposed by Loss and DiVincenzo \citep{loss-1998_PhysRevA.57.120}. In this way, the exchange interaction is maintained at a sweet spot with respect to charge fluctuations. Such a manipulation scheme is directly compatible in a quantum electronic circuit where the transfer of the single electron is performed with the help of SAWs \citep{hermelin-nature-2011}. In addition, the simplification in terms of dot tunability could have some impact on the scaling of the number of coherently controlled quantum dots.

\begin{figure}[h!]
\begin{center}
	\includegraphics[width=3in]{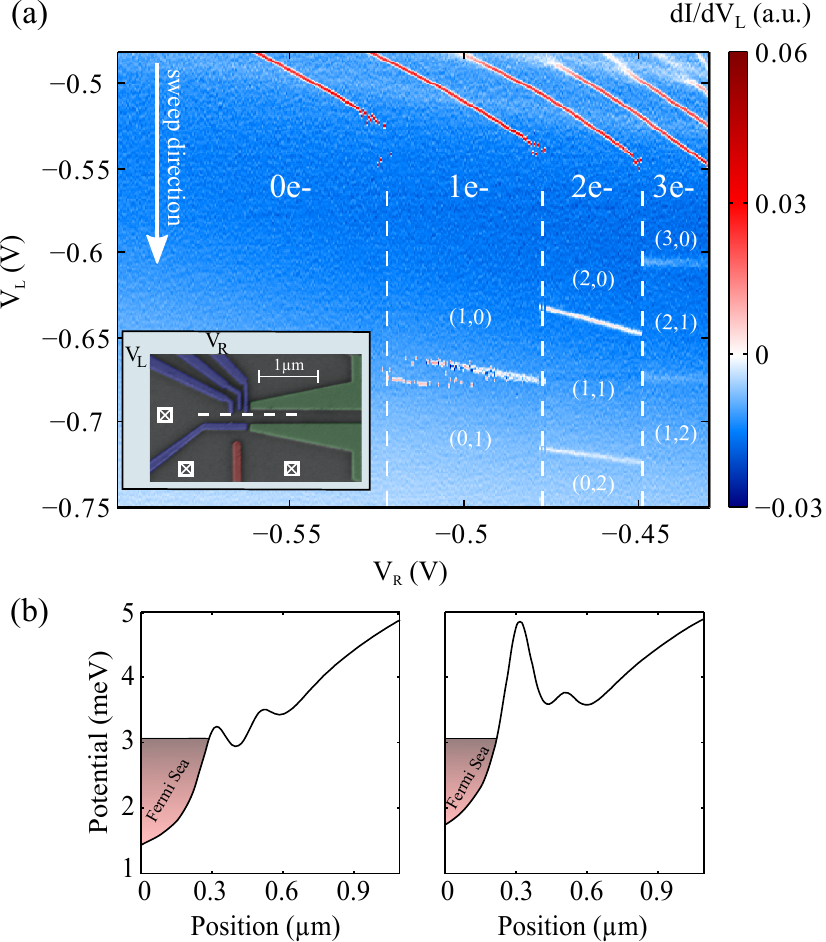}
\end{center}
\caption{(Color online) (a) Stability diagram of the charging quantum dot. The QPC current is averaged for $3$ ms per point, the voltage $V_L$ is swept from more positive to more negative values whereas $V_{R}$ is stepped. The signal is then numerically derived with respect to $V_L$ to highlight the changes in the QPC conductance. The number of charges loaded into a metastable position can be changed from 0 to 3 by varying $V_{R}$ . The label (i,j) refers to i electrons in the left part and j electrons in the right part of the metastable double dot. The (1,2)/(0,3) transition is hardly visible because of the large tunnel coupling between the dots at this position. (inset) SEM picture of the quantum dot system.
(b) Electrostatic calculation of the potential along the white dashed line (Fig. \ref{fig:1} (a) inset) for low (left) and high (right) negative voltage on gate $V_{L}$.
}
\label{fig:1}
\end{figure}

The quantum dot system is fabricated using a GaAs/AlGaAs heterostructure grown by molecular beam epitaxy with a 2DEG $\SI{100}{nm}$ below the surface. It is formed by local depletion of the 2DEG by means of metal Schottky gates deposited on the surface of the sample. The inset of Fig. \ref{fig:1}(a) shows a scanning electron microscopy (SEM) image of the sample used in the experiment \cite{hermelin-nature-2011}. An electrostatic calculation of the potential experienced by the electrons is presented in Fig. \ref{fig:1}(b). The left dot, called the charging dot, is the only one connected to the Fermi sea and permits the charging of the metastable double dot with electrons. The right one, called the channel dot, is realized with the two long gates and lies few meV above the Fermi energy. The voltage $V_R$ applied on the gate $R$ allows to tune the interdot tunnel coupling $t$. The voltage $V_L$ applied on the gate $L$ is used to close the barrier between the charging dot and the reservoir but also to change the detuning $\epsilon$ between the dots. Both gates are connected via low temperature home made bias tees to high bandwidth coaxial lines allowing GHz manipulation. The frequency range of the low-frequency section of the bias tees is DC-1 MHz permitting fast loading of the electrons. The charge state of the loading dot can be monitored using an on-chip electrometer (a quantum point contact (QPC) defined with the red gate in the inset of Fig. \ref{fig:1} (a)) with a 1~kHz detection bandwidth imposed by the room-temperature electronics. Measurements have been performed in a dilution refrigerator with a base temperature of $\SI{60}{mK}$. A $\SI{100}{mT}$ magnetic field is applied in the plane of the 2DEG.

\begin{figure} [h!]
\begin{center}
\includegraphics[width=3in]{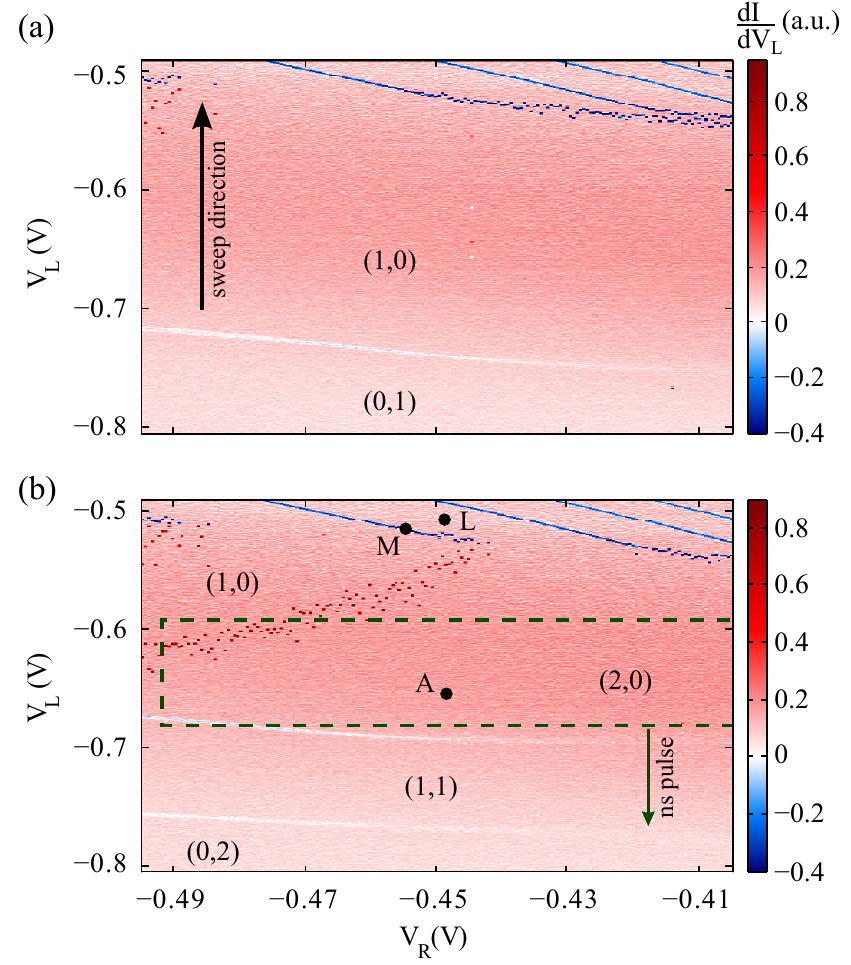}
\end{center}
\caption{Stability diagrams of the metastable double dot with 1 (a) and 2 (b) loaded electrons. $V_L$ is swept from negative to positive voltages. The white lines correspond to the charge degeneracy lines of the metastable double dot configuration. The stochastic red events, mostly visible in (b), correspond to tunneling events between the dots and the lead. The point L, M and A correspond respectively to the loading position for two electrons, the two-electron spin measurement position and a position with two electrons in the metastable charging dot.}
\label{fig:2}
\end{figure}

Figure \ref{fig:1} (a) shows a charge stability diagram of the charging quantum dot. For $V_{L}>-0.52$ V, the charging dot is well coupled to its reservoir and the charge degeneracy lines between $n+1$ and $n$ electrons are observed down to zero electrons. For $V_{L}<-0.52$ V, the charge degeneracy lines are disappearing which indicates that the exchange of electrons between the charging dot and the lead has been suppressed. Therefore the number of charges contained in the dot system remains fixed until the end of the sweep and can be changed with $V_R$. Deeper in this metastable region (lower part of Fig. \ref{fig:1} (a)) additional lines can be seen which are the results of tunneling to the channel quantum dot. Since electron exchange with the leads is forbidden, only the charge distribution between the dots can be varied: for the case of $n$ electrons contained in a metastable double dot, only $n$ charge degeneracy lines are expected. This leads to a drastic simplification of the obtained stability diagrams that could be useful for the control of multidot structures. 

\begin{figure}[h!]
\begin{center}
	\includegraphics[width=2.9in]{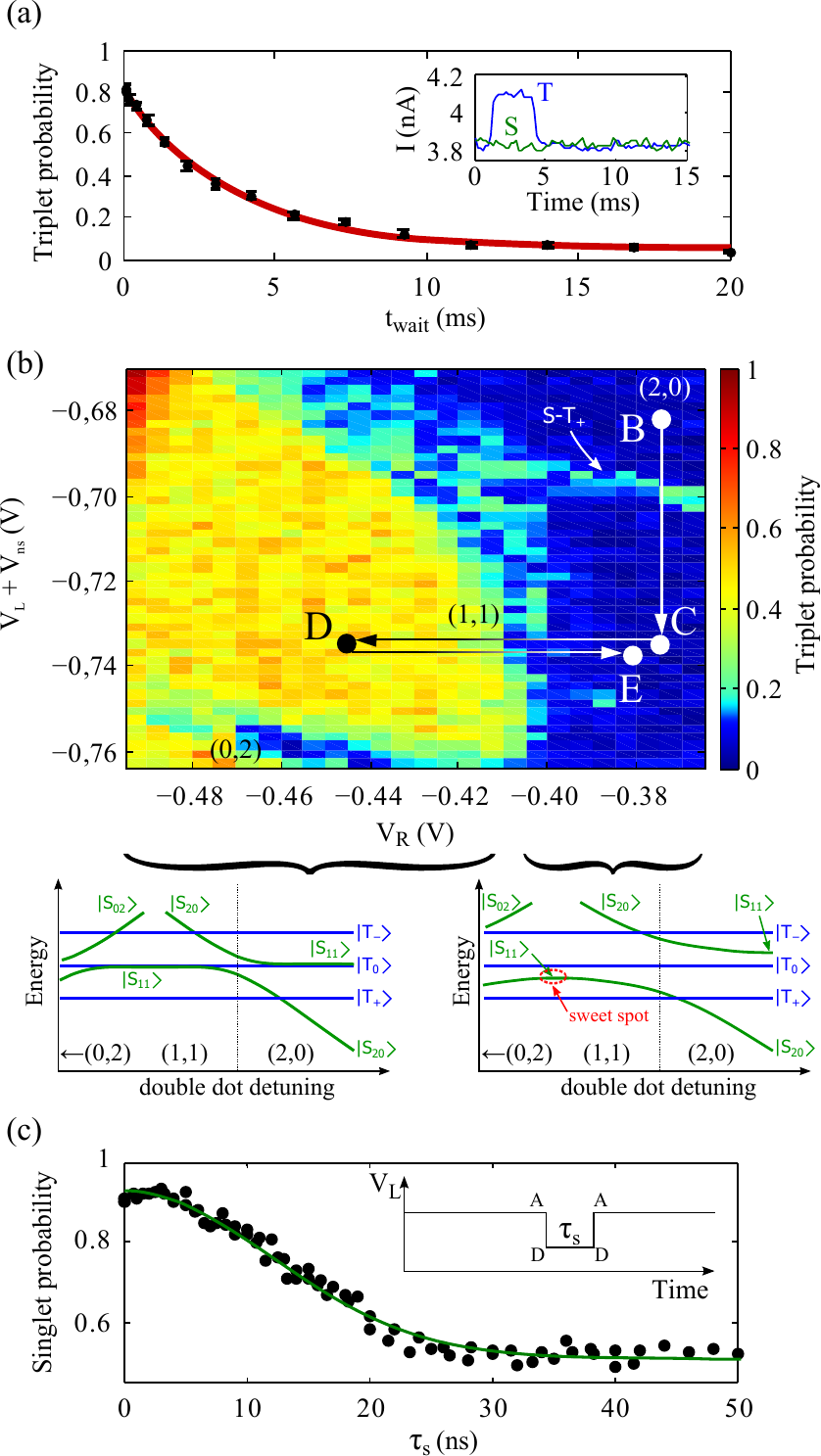}
\end{center}
\caption{(Color online) (a) Triplet to Singlet relaxation. Two electrons are loaded in the left dot at position $L$ on Fig.\ref{fig:2} (b) (mainly in triplet states), and the waiting time is varied to change the triplet proportion. The system is then brought to metastable position $A$ for 100 $\mu$s and finally to position $M$ to perform single shot energy selective spin measurement \cite{elzerman-2004, meunier-pss-2006}. This allows to distinguish single-shot the singlet and the triplet states with 80$\%$ fidelity. (Inset) Time dependence of the current through the electrometer when a singlet (green) or a triplet (blue) state is present in the dot. A tunneling event is only observed when the dot is occupied with a triplet state. (b) Spin mixing map of the two electron spin states (see text for details). Each data point is the result of an average over 200 single shot spin measurements. The corresponding energy diagrams of the two-electron spin states in the metastable double dot for low (left) and strong (right) tunnel coupling are shown below. $T_+$, $T_0$ and $T_-$ are the three triplet states in the (1,1) charge configuration. $S(0,2)$, $S(1,1)$ and $S(2,0)$ are the singlet state respectively in the (0,2), (1,1) and (2,0) charge configuration. (c) Measurement of the singlet probability by varying the time spent in the (1,1) configuration where singlet-triplet mixing occurs. Data is fitted with a Gaussian decay of $\SI{17}{ns}$. (Inset) Pulse sequence applied to $V_L$ at $V_R=-0.44$ V. 
}
\label{fig:3}
\end{figure}

In the metastable configuration, we are able to characterize the double dot with a fixed number of electrons over a wide range of gate voltages. It is indeed possible to load first the charging dot with the desired number of electrons and then rapidly promote them into the metastable position with a microsecond gate pulse. Finally, the system is scanned from that position to reconstruct a stability diagram of the metastable double dot. Figure \ref{fig:2} shows the observed stability diagram with the overall electron number fixed to one (two) and shows one (two) continuous interdot charge degeneracy lines. Contrary to the case where double dots are coupled to the leads, $t$ can be tuned over many orders of magnitude while keeping the number of electrons constant. Indeed, stochastic tunneling events are observed for the most negative values on $V_{R}$, giving a tunnel coupling smaller than the measurement bandwidth ($kHz$). For less negative $V_{R}$, the lines are broadened until being too wide to be seen, implying that the tunnel coupling overcomes the effect of temperature ($GHz$). The demonstrated control over $t$ is at the heart of the scheme initially proposed by Loss and DiVincenzo \citep{loss-1998_PhysRevA.57.120} to couple two electron spins efficiently via exchange interaction.

For large $t$, the exchange coupling $J$ is the dominant interaction and the singlet is a good eigenstate when one electron is present in each dot. For small $t$, $J$ becomes negligible with respect to the effective magnetic field gradient $\Delta B_z$ induced by the coupling to the nuclear spins of the heterostructure \cite{S-T-relaxation}. As a consequence, mixing between singlet and triplet states occurs. To infer the relative strength between $J$ and $\Delta B_z$, we probe where spin states are mixed in the gate voltage space. The protocol is as follows: two electrons are first injected in the charging dot and initialized in the singlet ground state by waiting longer than the relaxation time at position $L$; second, the electrons are brought on microsecond timescales to position $A$ ; third, we scan the system with a microsecond pulse over the dashed green rectangular region depicted on Fig. \ref{fig:2}(b) while adding a fast pulse of duration 50 ns and of amplitude 80 mV to the gate $V_L$; finally we proceed to the spin measurement in bringing the system to point $M$ in Fig. \ref{fig:2}(b) where energy selective spin read-out is performed (see Fig. \ref{fig:3} for details). The resulting spin mixing map is presented in Fig. \ref{fig:3} (b). In the (2,0) region, the system remains in the singlet state. Once the (2,0)-(1,1) charge transition is crossed, we observe spin mixing in the (1,1) charge region only for $V_{R}<-0.40$ V. This is consistent with the small tunneling region identified in Fig. \ref{fig:2} where $J$ is supposed to be smaller than $\Delta B_z$. We confirmed our interpretation by analyzing the typical timescale for mixing. The data are presented in Fig. \ref{fig:3} (c) and are characterized by a Gaussian decay of 17 ns, a timescale comparable to the one reported for double dots coupled with the leads \cite{S-T-relaxation}. For $V_{R}<-0.40$ V, $J$ becomes dominant with respect to $\Delta B_z$ and the system remains in the singlet state. No mixing is observed in this gate voltage region except for a thin line. It arises from the $S-T_+$ crossing \citep{Petta-science-2005}, consistent with our interpretation. We can confirm its nature  from the dependence of the line with the external magnetic field (data not shown). We therefore demonstrate that we can control $J$ with $t$ from kHz to several GHz in metastable double dots.

\begin{figure}[h!]
\begin{center}
	\includegraphics[width=2.9in]{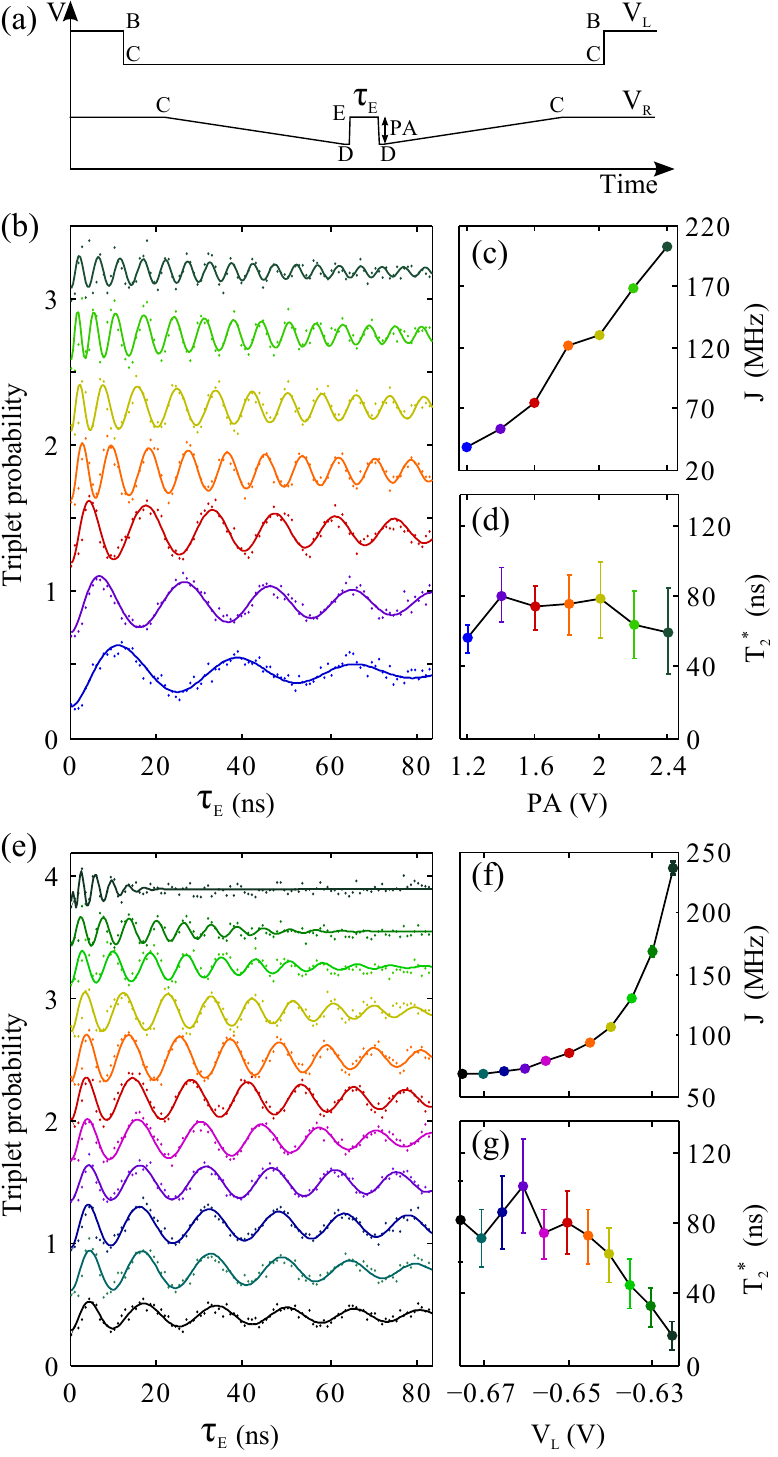}
\end{center}
\caption{(Color online) (a) Pulse sequence of the coherent exchange oscillations (see text for details). Coherent exchange oscillations: Triplet probability as a function of $\tau_E$ for different PA on $V_R$ with $V_L=\SI{-0.645}{V}$ (b) and for different $V_L$ at fixed PA$=\SI{1.6}{V}$ (e). Curves are vertically offset for clarity. Solid lines corresponds to a fit to the data with a damped oscillatory evolution. Extracted oscillation frequency $J$ as a function of PA (c) and $V_L$ (f). Extracted effective decoherence time $T_2^*$ as a function of PA (d) and $V_L$ (g).
}
\label{fig:4}
\end{figure}

To induce coherent flip-flops between electron spins, switching $J$ on and off with $t$ has the main advantage of keeping the system in a sweet spot $dJ/d\epsilon=0$ \cite{loss-1998_PhysRevA.57.120} (see Fig.\ref{fig:3} c). The coherent exchange pulse sequence is presented in Fig. \ref{fig:4} (a). A two-electron singlet state is initialized and brought to the metastable double dot in the (2,0) charge state (point B). A nanosecond pulse on $V_L$ brings the system to the (1,1) region with large $t$ (point C), where the singlet is still a good eigenstate. The system is then pulsed adiabatically to point D where $t$ is small and ends up in the eigenstate state of the Overhauser field $\ket{\uparrow \downarrow}$. Finally a fast pulseon $V_L$ of duration $\tau_E$ is applied to reach point E which projects this state back into the $S-T_0$ basis and allows rotation from $\ket{\uparrow \downarrow}$ to $\ket{\downarrow \uparrow}$ at a frequency $J/h$. The pulse amplitude (PA) defines the tunnel coupling $t_{p}$ during the coherent evolution. A mirror sequence followed by a spin measurement enables to recover the triplet probability after the rotation. The resulting coherent oscillations for different $t_p$ are presented in Fig. \ref{fig:4} (b). The contrast of the oscillations reaches only 50$\%$ which is attributed to imperfect preparation of the Zeeman eigenstates via adiabatic passage from the initial singlet state in the charging dot. As expected, the larger $t_p$ is, the larger $J$ will be. When fitting $J$ with a damped oscillatory model, we find an almost linear dependence of the oscillation frequency with PA. No significant change in $T^*_2$ is observed, which is consistent with the linear dependence of $J$ with PA on $V_R$. 

In Fig. \ref{fig:4}(c), we present a series of coherent exchange oscillations at fixed PA for different $V_L$ and therefore different $\epsilon$ between the two dots. In this way, we leave the sweet spot and the system becomes sensitive to the charge noise in $\epsilon$. Indeed, the closer to the (1,1)-(2,0) crossing the system is, the more important the influence of $\epsilon$ on $J$ and on $T^*_2$ should be. As expected, we observe a significant acceleration of $J$ as the system approaches the (1,1)-(2,0) crossing. This observation is consistent with the exponential $\epsilon$-dependence of $J$ reported previously \citep{Petta-science-2005, thalineauPRBCPhase}. It implies that the sensitivity $dJ/d\epsilon$ to noise in detuning also increases exponentially. This is consistent with the reduction of $T_2^*$ as the system approaches the (1,1)-(2,0) crossing and the drastic reduction of the observed number of oscillations. The reported behavior has to be compared with the previous situation where $T_2^*$ remains constant when $t_p$ is increased. Therefore our analysis demonstrates the advantage of the coherent exchange manipulation controlled by the tunneling. 

In conclusion, we have investigated the dynamics of two-electron spin states in a metastable double quantum dot. The extreme tunability of the dot system allows to map the mixing process between the two-electron spin states over a wide range of detuning and tunnel-coupling between the dots. It enabled us to perform coherent spin flip-flop between two electron spins in keeping the system in a sweet spot with respect to the detuning charge noise. This work demonstrates coherent manipulations compatible with fast and efficient transfer of a single electron with surface acoustic waves and paves the way towards the coherent control of multi-tunnel-coupled quantum dots.

\begin{acknowledgments}

We acknowledge technical support from the "Poles" of the Institut N\'eel as well as from Pierre Perrier. M.Y. acknowledges financial support by JSPS, Grant-in-Aid for Scientific Research A (No. 26247050) and Grant-in-Aid for Challenging Exploratory Research (No. 25610070). S. Tarucha acknowledges financial support by JSPS, Grant-in-Aid for Scientific Research S (No. 26220710), MEXT KAKENHHI "Quantum Cybernetics”, MEXT project for Developing Innovation Systems, and JST Strategic International Cooperative. A.L. and A.D.W. acknowledges expert help from Dirk Reuter and A. K. Rai and support of the BMBF  Q.com-H  16KIS0109, Mercur  Pr-2013-0001 and the DFH/UFA  CDFA-05-06. T.M. acknowledges financial support from ERC "QSPINMOTION".

\end{acknowledgments}

\end{document}